# Neutron diffraction and *ab initio* studies on the fully compensated ferrimagnetic characteristics of $Mn_2V_{1-x}Co_xGa$ Heusler alloys

P V Midhunlal[1], C Venkatesh[2], J Arout Chelvane[3], P D Babu[4] and N Harish Kumar[1]

1. Department of Physics, Indian Institute of Technology Madras, Chennai-600036, India.
2. Saha Institute of nuclear physics, Bidhannagar, Kolkata, West Bengal-700064.
3. Defense Metallurgical Research Laboratory, Kanchanbagh (PO), Hyderabad-500058, India.
4. UGC-DAE Consortium for Scientific Research, Mumbai Center, R-5 shed, BARC, Trombay, Mumbai-400085, India.

**Abstract:** Neutron diffraction and *ab initio* studies were carried out on $Mn_2V_{1-x}Co_xGa$ (x=0, 0.25, 0.5, 0.75, 1) Heusler alloys which exhibits high $T_C$ fully compensated ferrimagnetic characteristic for x=0.5. A combined analysis of neutron diffraction and *ab initio* calculations revealed the crystal structure and magnetic configuration which could not be determined from the X-ray diffraction and magnetic measurements. As reported earlier, Rietveld refinement of neutron diffraction data confirmed $L2_1$ structure for $Mn_2VGa$ and $X_a$ structure for $Mn_2CoGa$. The alloys with x=0.25 and 0.5 possess $L2_1$ structure with $Mn_{(C)}$-Co disorder. As the Co concentration reaches 0.75, a structural transition has been observed from disordered $L2_1$ to disordered $X_a$. Detailed *ab initio* studies also confirmed this structural transition. The reason for the magnetic moment compensation in $Mn_2(V_{1-x}Co_x)Ga$ was identified to be different from that of the earlier reported fully compensated ferrimagnet (MnCo)VGa. With the help of neutron diffraction and *ab initio* studies, it is identified that the disordered $L2_1$ structure with antiparallel coupling between the ferromagnetically aligned magnetic moments of ($Mn_{(A)}$-$Mn_{(C)}$) and (V-Co) atom pairs enables the compensation in $Mn_2V_{1-x}Co_xGa$.

# 1. Introduction

Half-metallic fully compensated ferrimagnets (HMFCFi) with high Curie temperatures are potential candidates for spintronic applications especially for the spin-transfer torque based random access memory (STT-RAM) [1]. They possess 100 % spin polarization at the Fermi level along with zero net magnetization and produces no stray magnetic field which increases the efficiency of spintronic devices. Even though HMFCFi possesses zero magnetic moment macroscopically, they significantly differ from conventional antiferromagnets. Unlike the conventional antiferromagnets, zero magnetization occurs in HMFCFi due to the magnetic moment compensation arising from the different sub-lattices which are chemically or crystallographically inequivalent [2]. There are reports which discuss the possibility of having zero moment half-metallicity in different kinds of materials such as Heusler alloys [3], dilute magnetic semiconductors [4], double perovskites [5] and transition metal pnictides [6]. Among these, the most favourable candidate is Heusler alloys which are ternary intermetallics having general formula $X_2YZ$ for full-Heusler alloys and XYZ for half-Heusler alloys (where X, Y are transition metal and Z is an sp valent atom). The band structure and thereby the physical properties of these alloys can be tuned by modifying the chemical composition through the substitution of certain elements at specific sites. The Slater-Pauling rule (S-P rule), i.e., $M_t=Z_t-24$ (for full-Heusler alloys), which relates the total magnetic moment per formula unit ($M_t$) to the total number of valence electrons ($Z_t$) of half-metallic Heusler alloys can be utilized as an important tool to engineer new materials with improved physical properties like zero moment half-metallicity. The rule predicts half-metallic zero magnetic moment state in full-Heusler alloys having 24 valence electrons [7]. Galanakis I *et al.* have shown that half-metallic fully compensated ferrimagnetism is achievable in well-known half-metallic $Mn_2VZ$ (Z=Al,Si) Heusler alloys by substituting Co at the Mn site [8]. The alloys (MnCo)VAl and $(Mn_{1.5}Co_{0.5})$VSi having 24 valence electrons are predicted to be HMFCFi. Based on their theoretical prediction, experimental investigations have been reported on $(Mn_{2-x}Co_x)$VZ (Z=Ga,Al) alloys [9,10]. This is interesting since the parent alloys $Mn_2VZ$ (Z=Ga,Al) are reported to have half-metallic characteristics [11,12]. Even though the results showed that the magnetic moment is compensating in these Co substituted systems, the Curie temperature decreases below room temperature with increase in Co concentration for both the alloy series which is not favourable for devise applications. On the other hand, our recent experimental investigation showed that the Co substitution at the V site has resulted in total magnetic moment compensation without significant change in Curie temperatures in the ferrimagnetic $Mn_2VZ$ (Z=Ga,Al) Heusler alloys [13]. The observed Curie temperatures were well above the room temperature (> 650 K) for $Mn_2V_{1-x}Co_xZ$ (Z=Ga,Al; x=0,0.25,0.5,0.75,1) alloys which were in contrast to the earlier reports on Co substitution at the Mn site. The observed magnetic moment and $T_C$ were 0.10 $\mu_B$/f.u. and 706 K for $Mn_2V_{0.5}Co_{0.5}Ga$ and 0.06 $\mu_B$/f.u. and 659 K for $Mn_2V_{0.5}Co_{0.5}Al$ respectively. The possible reason for the magnetic moment compensation was attributed to the antiparallel coupling between the ferromagnetically aligned moments of ($Mn_{(A)}$-$Mn_{(C)}$) and (V-Co) atom pairs. But there was no experimental or theoretical evidence for this proposed magnetic configuration. Thus in order to determine the crystal structure and magnetic configuration of the interesting alloys $Mn_2V_{1-x}Co_xGa$ (x=0,0.25,0.5,0.75,1), neutron diffraction measurement and density functional theory (DFT) based *ab initio* calculations were carried out. The crystal structure and the magnetic configuration were identified by comparing the neutron diffraction (ND) results with that of the *ab initio* calculations.

## 2. Experimental & Computational details

The details on sample preparation, discussions on X-ray diffraction and magnetic measurements can be found in our earlier report [13]. Room temperature neutron diffraction (ND) measurements were carried out at a wavelength of 1.48 Å using the position-sensitive detector based focusing crystal diffractometer PD-3 installed by the UGC-DAE CSR Mumbai Centre at the Dhruva reactor, Trombay [14]. Spin-polarized band structure calculations were carried out for the $Mn_2V_{1-x}Co_xGa$ (x= 0, 0.25, 0.5, 0.75, 1) Heusler alloys using full-potential linearized augmented plane wave method (FP-LAPW) implemented in WIEN2k code [15]. Pedrew-Burke-Ernzerhof parameterization scheme was used for the generalized gradient approximation (GGA) exchange-correlation potential [16]. The calculations were carried out for the 16 atoms supercell. The k-space integration was carried out by the tetrahedron method [17] by taking 150 k points in the irreducible Brillouin zone. The plane wave cut off parameter $R_{MT} \times K_{max}$ was taken as 9. $R_{MT} \times K_{max}$ is the product of the smallest atomic sphere radius times the largest K-vector of the plane wave expansion of the wave function which determines the size of the basis set and thereby the accuracy of the calculation. The angular momentum truncation parameter $l_{max}$ was set to 9. The energy convergence (-ec) and charge convergence (-cc) criterion was set to $10^{-4}$ Ry and $10^{-4}$ e respectively.

$X_2YZ$ Heusler alloys (X, Y are transition metal atom and Z is sp valent atom) crystallize in the cubic $Cu_2MnAl$ type structure (Space group: 225, $L2_1$, $Fm\bar{3}m$) having four interpenetrating fcc sub-lattices namely A, B, C and D with coordinates (0,0,0), $(\frac{1}{4},\frac{1}{4},\frac{1}{4})$, $(\frac{2}{4},\frac{2}{4},\frac{2}{4})$ and $(\frac{3}{4},\frac{3}{4},\frac{3}{4})$ respectively. X atoms are equally occupied in the A and C sub-lattices. The Y and Z atoms occupy the B and D sub-lattices respectively. Apart from the usual $Cu_2MnAl$ structure, Heusler alloys are found to crystallize in $Hg_2CuTi$ type structure also, commonly known as inverse Heusler structure (space group: 216, $X_a$, $F\bar{4}3m$). Unlike the $L2_1$ alloys, the valance of X transition-metal atom is smaller than that of the Y transition-metal atom for the $X_a$ alloys and the atoms in the B and C sub-lattices will be swapped as shown in figure 1(a)&(b). In the case of $Mn_2V_{1-x}Co_xGa$, the atomic ordering along cube diagonal would be $Mn_{(A)}$- $(V/Co)_{(B)}$-$Mn_{(C)}$-$Ga_{(D)}$ in the $L2_1$ structure and $Mn_{(A)}$-$Mn_{(B)}$-$(V/Co)_{(C)}$-$Ga_{(D)}$ in the $X_a$ structure. So the structural refinement of ND pattern and *ab initio* calculations were carried out by considering both $L2_1$ and $X_a$ crystal structures.

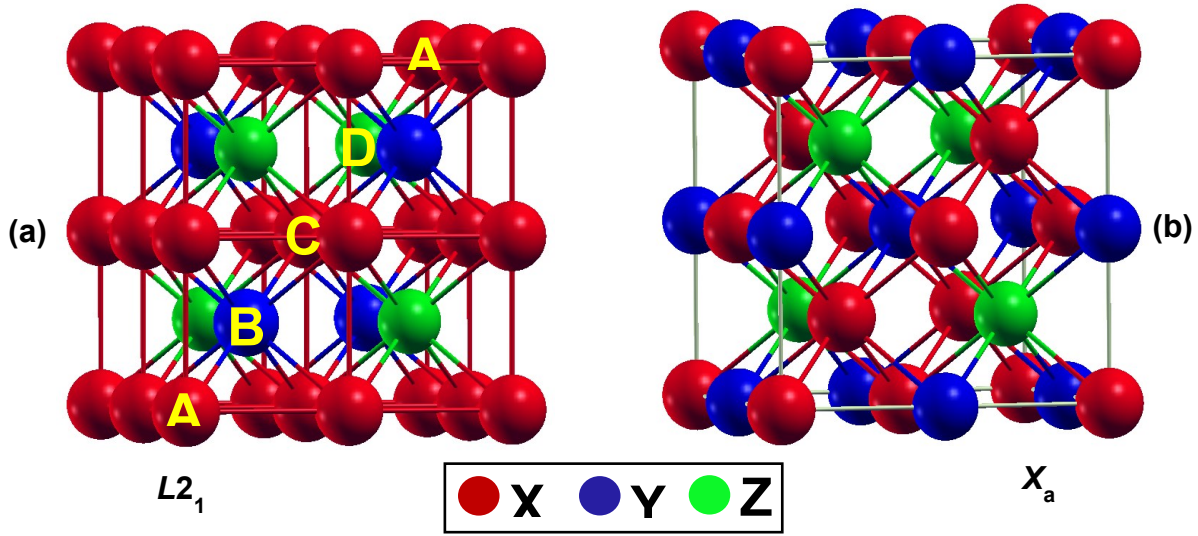

**Figure 1**: (a) $L2_1$ crystal structure with $X_{(A)}$-$Y_{(B)}$-$X_{(C)}$-$Z_{(D)}$ atomic configuration along the cube diagonal (b) $X_a$ crystal structure with $X_{(A)}$-$X_{(B)}$-$Y_{(C)}$-$Z_{(D)}$ atomic configuration. A, B, C and D represents the four interpenetrating fcc lattices.

# 3. Results & discussion

## 3.1 Neutron diffraction studies on $Mn_2V_{1-x}Co_xGa$

Earlier neutron diffraction studies showed that $Mn_2VGa$ alloy were in the cubic $L2_1$ structure [18]. So in our previous report on the structural and magnetic properties of the $Mn_2V_{1-x}Co_xGa$ (x=0,0.25,0.5,0.75,1) alloys, Rietveld refinement of X-ray diffraction (XRD) patterns was carried out by assuming the $L2_1$ structure. At the same time, the report by Liu *et al.* shows that the other end member $Mn_2CoGa$ crystallize in the inverse ($X_a$) crystal structure [19]. So the crystal structure of substituted alloys could be either $L2_1$ or $X_a$. This was not taken into consideration in our previous report. So in order to remove the uncertainty in the crystal structure, we have carried out the refinement by considering the $X_a$ structure also. The Rietveld refinement carried out on $Mn_2V_{1-x}Co_xGa$ alloys showed that the X-ray diffraction patterns were fitting well for both $L2_1$ and $X_a$ structural models resulting in the similar lattice parameter and $\chi^2$ values. This makes it difficult to distinguish the crystal structure of these Co substituted alloys through XRD analysis. To illustrate this problem in a better manner, XRD patterns of $Mn_2V_{1-x}Co_xGa$ (x=0,0.25,0.5,0.75,1) alloys in both $L2_1$ and $X_a$ crystal structures have been simulated using FullProf software [20] as displayed in figure 2(a)&(b). It is clear that there are no noticeable differences in the simulated $L2_1$ and $X_a$ X-ray diffraction patterns. Our earlier experimental report showed similar X-ray diffraction patterns without having any order dependent superlattice reflections (111) and (200). The absence of superlattice reflections was attributed to the similar atomic scattering factors of X-ray for the elements Mn, V, Ga and Co. The presence of atomic anti-site disorder was also unavoidable. The ambiguity in the crystal structure could be resolved by doing neutron diffraction (ND) measurements since the coherent scattering length for individual elements differs significantly (Mn= -3.73 fm; V= -0.38 fm; Ga= 7.288 fm; Co= 2.49 fm). In order to check this, the neutron diffraction patterns of the alloys were simulated as shown in figure 3(a)&(b). For simulating the X-ray and neutron diffraction patterns, earlier reported experimental lattice parameter was used. Unlike the X-ray diffraction patterns, the simulated ND patterns show that except the (333) peak, all other peaks are paired in both $L2_1$ and $X_a$ crystal structure. $L2_1$ and $X_a$ crystal structures are easily distinguishable by looking at the relative intensities of these paired peaks which were not the case with the X-ray diffraction patterns. For the $L2_1$ structure, the second peak in each pair is more intense than the first peak and for the $X_a$ structure, the intensity of the first peak is higher than that of the second one. Another noticeable difference between the two simulated ND pattern is that the (333) peak in $X_a$ structure is more intense than that in the $L2_1$ structure. Here a striking difference between the simulated XRD and the ND patterns has to be mentioned. The visible peaks such as (220), (400), (422), (440) and (620) in the XRD patterns (indexed in green) are having negligible intensity in the ND patterns. At the same time, the low intense peaks such as (111), (200), (311), (222), (331), (420) etc… in the XRD patterns (indexed in black) are pronounced in the ND patterns. Thus the simulated ND patterns clearly show the difference between $L2_1$ and $X_a$ crystal structure.

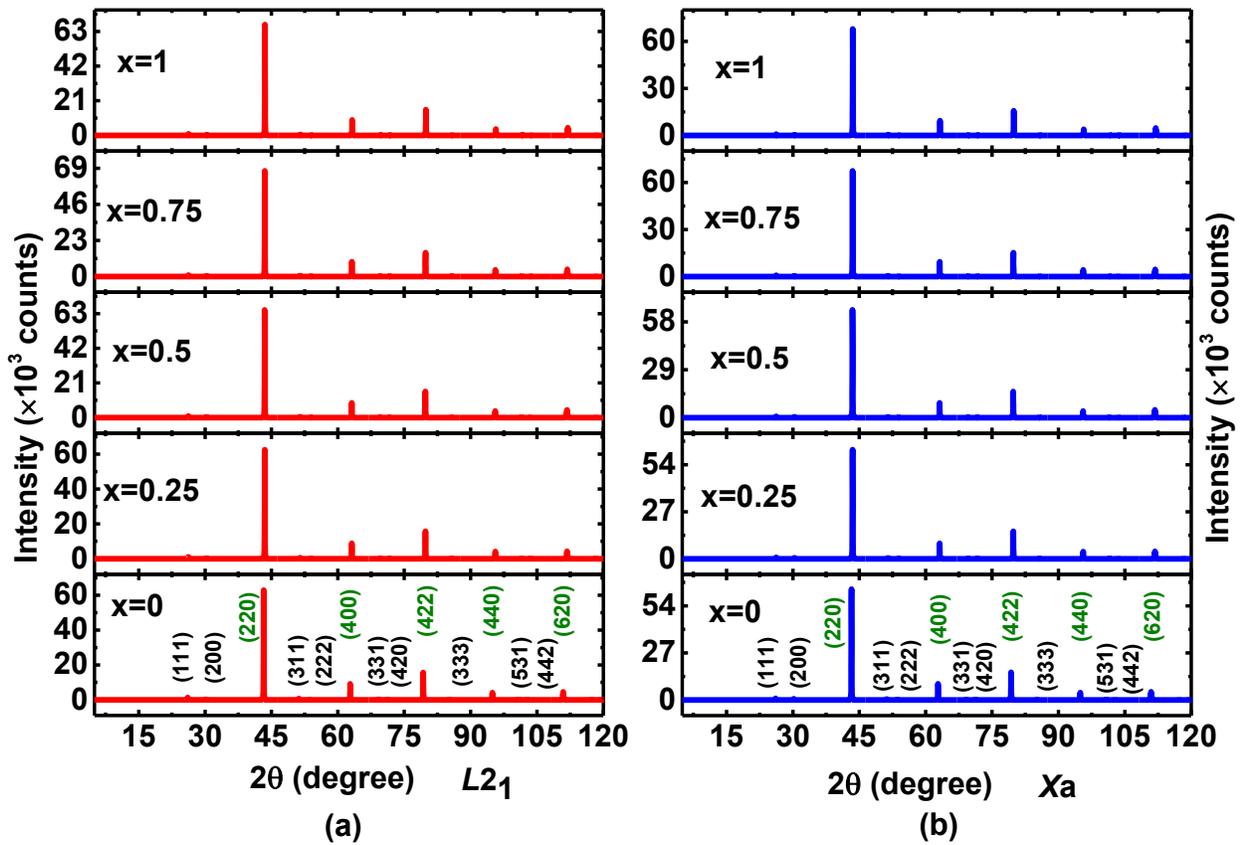

**Figure 2**: Simulated XRD patterns of $Mn_2V_{1-x}Co_xGa$ alloys in the (a) $L2_1$ and (b) $X_a$ crystal structure.

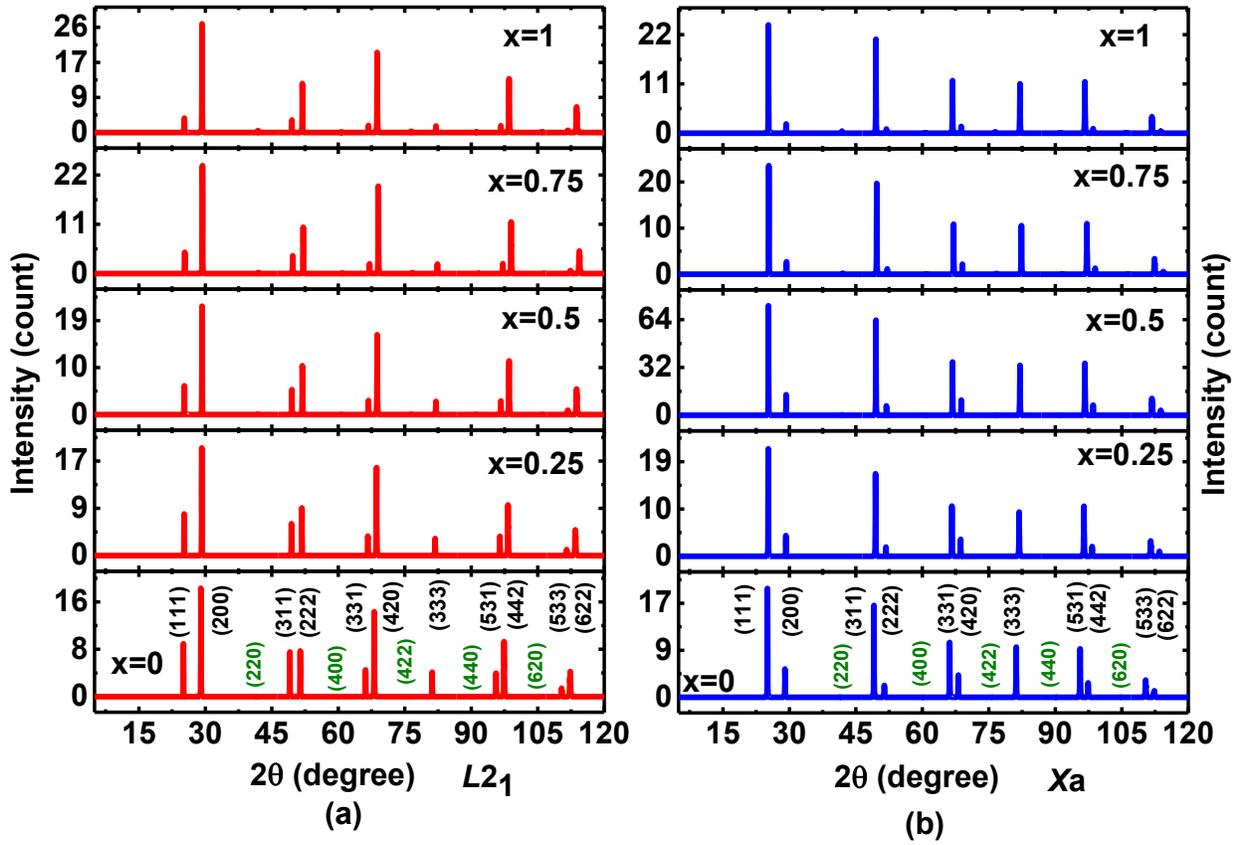

**Figure 3**: Simulated neutron diffraction patterns of $Mn_2V_{1-x}Co_xGa$ alloys in the (a) $L2_1$ and (b) $X_a$ crystal structure.

As the ND patterns would distinguish the $L2_1$ and $X_a$ crystal structure, we have recorded the ND patterns of $Mn_2V_{1-x}Co_xGa$ alloys at room temperature which is shown in figure 4(a). Before discussing the Rietveld refined patters as shown in figure 4(b), some distinguishable features can be observed from the unrefined ND patterns. Comparison of the simulated $L2_1$ and $X_a$ ND patterns with experimental data reveals that $Mn_2VGa$ crystallizes in $L2_1$ structure and $Mn_2CoGa$ crystallizes in $X_a$ structure. The relative intensity of the nearby peaks is in good agreement with the simulated patterns. It is evident that the order dependent superlattice reflections (111) and (200) are present with a huge intensity which was absent in the XRD patterns. So the reported absence of superlattice reflections in the XRD patterns of $Mn_2V_{1-x}Co_xGa$ alloys can be attributed to the similar atomic scattering factors of X-ray for the elements Mn, V, Ga and Co which are in the same period and not due to any atomic antisite disorder.

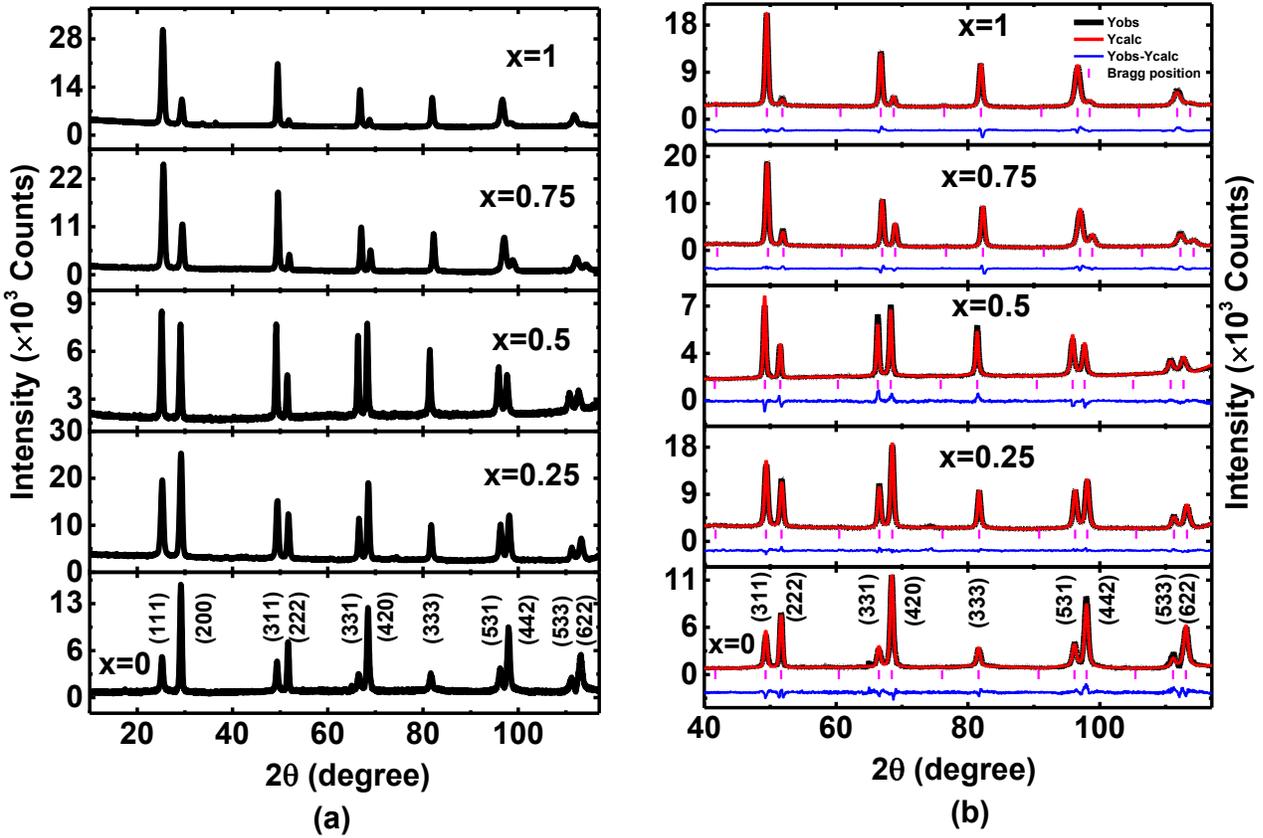

**Figure 4**: (a) Neutron diffraction patterns of $Mn_2V_{1-x}Co_xGa$ alloys recorded at room temperature. (b) Nuclear refinement of the data for $2\theta>40$.

For the Co substitution with x=0.25, some changes in the ND pattern can be observed compared to the parent $Mn_2VGa$. The relative intensity of the (111) peak has been increased and the intensity of the (311) peak is more compared to the (222) peak. Apart from these intensity changes the overall pattern resembles that of the $L2_1$. The changes in intensity suggest the possibility of having an atomic disorder which will be discussed later. When the Co concentration is increased to x=0.5, more changes in the peak intensities were observed. The relative intensity of peaks in the pairs (111) & (200), (311) & (222), (531) & (442) and the intensity of the single peak (333) resembles the $X_a$ structure and the pairs (331) & (420), (533) & (622) resembles the $L2_1$ structure. Another interesting point to note is that the intensity mismatch between the peaks in each pair is less compare to the simulated patterns. The above points indicate that the crystal structure of Mn$_2$V$_{0.5}$Co$_{0.5}$Ga alloy could be either $X_a$ or $L2_1$ with atomic antisite disorder. For the Co concentration with x=0.75, the ND pattern is similar to that of Mn$_2$CoGa which is $X_a$. This implies that there could be a structural transition from $L2_1$ to $X_a$ as the Co concentration increases which should be confirmed by doing the Rietveld refinement. By keeping these things in mind, we tried to do the Rietveld refinement for the ND pattern by using the FullProf software. Since Mn$_2$V$_{1-x}$Co$_x$Ga alloys are ferrimagnetic at room temperature (T$_C$ > 650 K), nuclear refinement was carried out by considering the scattering angle (2θ) above 40˚ at the beginning of the refinement. The refined ND patterns are shown in figure 4(b). The refinement showed that the Mn$_2$VGa alloy crystallizes in the $L2_1$ structure with 6 % atomic antisite disorder between V and Ga. For x=0.25 and 0.5, different structural models such as $L2_1$, $X_a$, $L2_1$ + (various disorders), $X_a$ + (various disorders) have been considered. For x=0.25, the best fit was obtained only for $L2_1$ + (Mn$_{(C)}$-Co disorder). For the alloy with x=0.5, the structural models $L2_1$ + (Mn$_{(C)}$-Co disorder) and $X_a$ + (Mn$_{(B)}$-Co disorder) have resulted in identical fit to the pattern. The ambiguity in crystal structure for this alloy has been removed with the help of *ab initio* calculations which is discussed in the next section. This calculation identified the correct structure to be $L2_1$+(Mn$_{(C)}$-Co) disorder. For the alloy with x=0.75, $X_a$ + (Mn$_{(B)}$-Co disorder) structural model has given the best fit. Mn$_2$CoGa was found to crystallize in the $X_a$ structure with 5 % atomic antisite disorder between Mn$_{(B)}$ and Co. The considered structural models, the observed disorders, lattice parameter and the respective $\chi^2$ values are shown in Table 1. It was evident from the refinement that the majority of the substituted Co has gone to the Mn site and the corresponding Mn atoms have gone to the V site for the alloys with x=0.25 and 0.5. The antisite Co content was reduced from 89 % to 5 % as the Co concentration increases from x=0.25 to 1. The obtained crystal structure after ND pattern refinement is in agreement with the preliminary observations. Interesting point is that there is a gradual structural transition from $L2_1$ to $X_a$ as the Co concentration increases from x=0 to 1 which was not evident from the XRD analysis.

| x | Fitted structural models | Co disorder (%) | Lattice parameter (Å) | $\chi^2$ |
|---|---|---|---|---|
| 0 | $L2_1$ + (V-Ga disorder) | 6 | 5.895 | 11 |
| 0.25 | $L2_1$ + (Mn$_{(C)}$-Co disorder) | 89 | 5.878 | 4.6 |
| 0.5 | $L2_1$ + (Mn$_{(C)}$-Co disorder) or $X$a + (Mn$_{(B)}$-Co disorder) | 86 71 | 5.862 5.866 | 5.7 4.6 |
| 0.75 | $X$a + (Mn$_{(B)}$-Co disorder) | 25 | 5.844 | 8.1 |
| 1 | $X$a + (Mn$_{(B)}$-Co disorder) | 5 | 5.862 | 4.1 |

**Table 1**: The considered structural models, the observed disorders, lattice parameter and the respective $\chi^2$ values for neutron diffraction refinement of Mn$_2$V$_{1-x}$Co$_x$Ga alloys.

Having identified the crystal structure of $Mn_2V_{1-x}Co_xGa$, the next step is to do the refinement by considering the magnetic contribution to the neutron diffraction pattern by taking the whole scattering angle 5-120°. The magnetic refinement showed that the ferromagnetically aligned Mn atoms in the A and C sublattices couples ferrimagnetically with V atoms in $Mn_2VGa$. The observed magnetic moments were -1.28 $\mu_B$ for $Mn_{(A)}$ & $Mn_{(C)}$ and 0.71 $\mu_B$ for V which sums up to -1.85 $\mu_B$/f.u. For $Mn_2CoGa$, refinement showed that the Mn atoms in the A and B sublattices are aligned ferrimagnetically with a moment of -1.82 $\mu_B$ for $Mn_{(A)}$, 3.20 $\mu_B$ for $Mn_{(B)}$ and 0.52 $\mu_B$ for Co which sums up to 1.9 $\mu_B$/f.u. For the substituted alloys (x=0.25, 0.5, 0.75), the magnetic phase refinement of ND data was not possible due to certain constraints. Since the disordered $L2_1$ and $X_a$ structures have 6 magnetic atoms per formula unit ($Mn_{(A)}$, $Mn_{(C/B)}$, $V_{(B/C)}$, $Co_{(B/C)}$, interchanged Mn and Co), fixing the individual moment became difficult. For example, in the case of x=0.25, As the 'C' sublattice is occupied by Mn and Co atoms and the 'B' sublattice is occupied by V, Co and Mn atoms, it is difficult to figure out the individual atomic contribution to the total neutron scattering amplitude from these sub-lattices. Another difficulty is that, due to the ferrimagnetic nature, magnetic reflections were overlapped with the nuclear reflection in the diffraction patterns and very small magnetic contribution was observed in the low angle (111) and (200) superlattice reflections only. This adds to the difficulty of magnetic phase refinement as the goodness of the fit has to be monitored by observing these two peaks alone. So in order to identify the magnetic configurations and to confirm the crystal structure, density functional theory based (DFT) *ab initio* calculations were carried out. The following section discusses the *ab initio* calculations on $Mn_2V_{1-x}Co_xGa$ alloys. Conclusions on the crystal structure and the magnetic configurations are given on the basis of a combined analysis of ND and *ab initio* calculations.

## 3.2 *Ab initio* studies on $Mn_2V_{1-x}Co_xGa$ (x=0, 0.25, 0.5, 0.75 and 1)

Spin-polarized *ab initio* calculations were carried out for $L2_1$ and $X_a$ crystal structures for each x value. For each structure, energy minimization with respect to lattice parameter (volume) has been carried out and the lattice parameter corresponding to the minimum energy was used for the self-consistent field (scf) calculations. Experimental lattice parameter for $Mn_2V_{1-x}Co_xGa$ was used for the energy minimization. The Figure 5(a)-(e) shows the energy minimization with respect to the volume for $Mn_2V_{1-x}Co_xGa$ alloys in $L2_1$, $X_a$ and disordered structures.

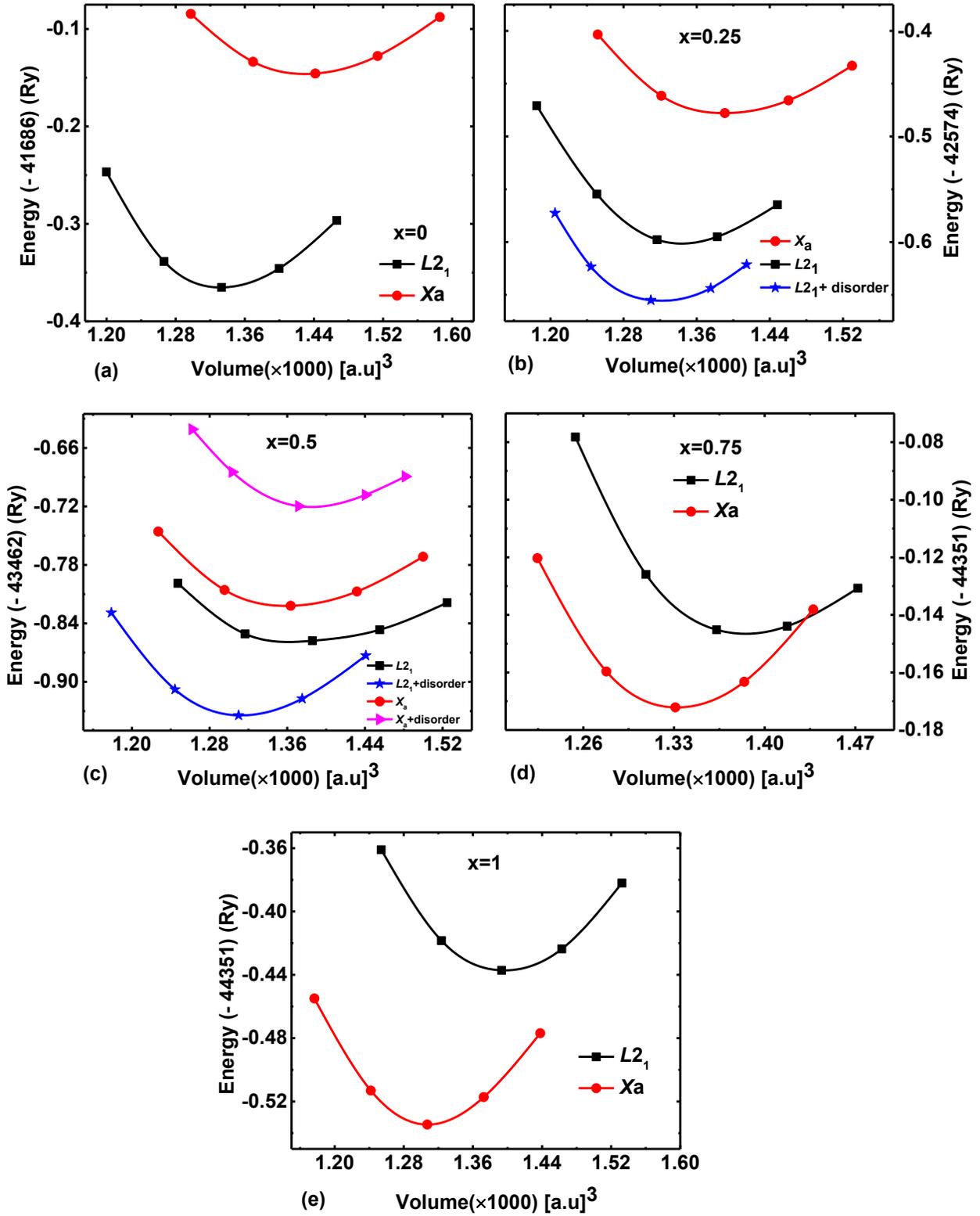

**Figure 5 (a)-(d)**: Energy minimization with respect to the volume for $Mn_2V_{1-x}Co_xGa$ alloys in $L2_1$, $X_a$ and disordered structures.

From the energy minimization, it is clear that the alloy with x=0 has minimum energy in the $L2_1$ crystal structure. ND analysis indicated that for the alloy with x=0.25, most of the substituted Co atoms have gone to the $Mn_{(C)}$ site and equivalent $Mn_{(C)}$ atoms have gone to the V site in the $L2_1$ structure. So along with the $L2_1$ and $X_a$ structures, $L2_1$+($Mn_{(C)}$-Co) disordered structure is also considered for the energy minimization. The stoichiometry used for the calculation is $(M_{1.75}Co_{0.25})(V_{0.75}Mn_{0.25})Ga$ which still has the total valence electron number of 23 and thereby a moment of -1 $\mu_B$/ f.u. as per the S-P rule. As shown in figure 5(b), $L2_1$+($Mn_{(C)}$-Co) disordered structure has exhibited lower energy which is in agreement with the ND result. Similarly, ND refinement suggested either a $L2_1$+($Mn_{(C)}$-Co) disordered structure or $X_a$+($Mn_{(B)}$-Co) disordered structure for x=0.5. So the energy minimization was carried out by considering these disordered structures also. The minimization curves clearly show that the $L2_1$+($Mn_{(C)}$-Co) disordered structure has the minimum energy compared to other structures. The stoichiometry used for the calculation is $(M_{1.5}Co_{0.5})(V_{0.5}Mn_{0.5})Ga$ which still has the total valence electron number of 24 and thereby a moment of 0 $\mu_B$/ f.u. as per the S-P rule. For x=0.75 and 1, $X_a$ crystal structure has the minimum energy compared to the $L2_1$ structure. This clearly shows a crossover of crystal structure from $L2_1$ to $X_a$ as the Co concentration increases which are in agreement with the ND results. Table 2 shows the calculated lattice parameters which are compared to that of the values reported in other experimental and *ab initio* studies. Table 3 shows the total and atom-resolved magnetic moments and spin polarization values of $Mn_2V_{1-x}Co_xGa$. The expected magnetic moments from S-P rule are -2, -1, 0, 1, and 2 $\mu_B$/f.u. and the reported experimental values in our earlier report are 1.80, 0.97, 0.10, 1.11 and 2.05 $\mu_B$/f.u for x=0, 0.25, 0.5, 0.75 and 1 respectively. The calculated total moments also follows the experimental trend of magnetic moment compensation as the Co concentration reaches x=0.5. For the x=0 alloy ($Mn_2VGa$), the calculated total magnetic moment in the lower energy $L2_1$ structure is -1.98 $\mu_B$/f.u. which is close to the S-P value.

| x | Crystal structure | Lattice parameter (Å) | | |
|---|---|---|---|---|
| | | Present Calculation | Other Reports | |
| | | | Cal. | Exp. |
| 0 | $L2_1$ | 5.824 | 5.820 [21] | 5.914 [13] |
| | | | 5.685 [22] | 5.905 [11] |
| | | | 5.808 [23] | 6.095 [24] |
| 0.25 | $L2_1$+disorder | 5.805 | - | 5.874 [13] |
| 0.5 | $L2_1$+disorder | 5.796 | - | 5.878 [13] |
| 0.75 | $X_a$ | 5.824 | - | 5.881 [13] |
| 1 | $X_a$ | 5.789 | 5.78 [21] | 5.877 [13] |
| | | | 5.61 [25] | 5.862 [19] |
| | | | | 5.869 [26] |

**Table 2**: Crystal structure and the corresponding calculated lattice parameters of $Mn_2V_{1-x}Co_xGa$ alloys. Calculated (Cal) and experimental (exp) lattice parameters reported by other researchers are also shown for the comparison.

| x | Crystal structure | Mag. Config. | atom resolved magnetic moments ($\mu_B$) | | | | Total moment (Cal) $\mu_B$/f.u. | Total moment (S-P rule) $\mu_B$/f.u. | Total moment (Exp)[13] $\mu_B$/f.u. | P (%) |
|---|---|---|---|---|---|---|---|---|---|---|
| | | | $Mn_{(A)}$ | $Mn_{(B/C)}$ | V | Co | | | | |
| | $L2_1$ | ddu | -1.45 | -1.45 | 0.82 | - | -1.98 | -2 | 1.80 | 94 |
| 0.25 | $L2_1$+disorder | dduu | -1.32 | -1.53 | 0.73  2.90 (Mn) | 0.48 | -0.99 | -1 | 0.97 | 99 |
| 0.5 | $L2_1$+disorder | dduu | -1.35 | -1.31 | 0.65  2.82 (Mn) | 0.66 | 0 | 0 | 0.10 | 99 |
| 0.75 | $X_a$ | dudu | -1.91 | 2.69 | -1.43 | 0.94 | 0.99 | 1 | 1.11 | 98 |
| 1 | $X_a$ | duu | -1.82 | 2.85 | - | 0.99 | 2.00 | 2 | 2.05 | 99 |

**Table 3**: Identified crystal structure, calculated (cal) atom-resolved and total magnetic moments and spin polarization values of $Mn_2V_{1-x}Co_xGa$. Experimental (exp) magnetic moments are also shown for the camparison. The antisite Mn moment is shown near to the V moment in the case of x=0.25 and 0.5. For representing the magnetic configuration of $Mn_{(A)}$ $Mn_{(B/C)}$V Co, upward and downward directions are denoted as 'u' and 'd' respectively.

The obtained moment value is in agreement with the earlier report [18]. The Mn moments in the A and C sublattices are found to be in ferromagnetic alignment which is common in the case of Mn-based Heusler alloys in the $L2_1$ structure. For $Mn_2CoGa$ (x=1), Mn moments in the A and B sublattices are found to be in antiparallel alignment which is observed in the case of Mn-based Heusler alloys in the $X_a$ structure. This is also in agreement with the earlier report [27]. For x=0.25 and 0.5, $L2_1$ structure with $Mn_{(C)}$-Co disorder is the minimum energy state having a total moment of -0.99 and 0 $\mu_B$/f.u. respectively. This is close to the experimental and S-P value (shown in Table 3). For x=0.75 also, the calculated moment is close to the experimental value. The obtained magnetic configuration from the *ab initio* calculation for the alloy with x=0.5 is similar to the assumption made for explaining the magnetic moment compensation in our previous report. $L2_1$ structure with **dduu** magnetic configuration was assumed for $Mn_{(A)}$, $Mn_{(C)}$, $V_{(B)}$, $Co_{(B)}$ atoms (here **u** implies upward moment direction and **d** implies downward moment direction). The present calculation also showed **dduu** configuration irrespective of the disorder. In addition to this, the disordered Mn atom at the V site has aligned in the upward direction. This implies that the $L2_1$ structure and ferromagnetic coupling of Mn atoms in the A and C sublattices of $Mn_2VGa$ have not altered with Co substitution up to x=0.5. Also, the substituted Co and V atoms couples ferromagnetically for x up to 0.5. In short, the disordered $L2_1$ structure with the ferrimagnetic coupling between the ferromagnetically aligned ($Mn_{(A)}$-$Mn_{(C)}$) and (V-Co) atom pairs enables the magnetic moment compensation in $Mn_2V_{1-x}Co_xGa$. Figure 6 shows the variation of atomic moments with the Co concentration. For the Co-substituted alloys, the $Mn_{(B/C)}$ and Co moment increases and the $Mn_{(A)}$ and V moment decrease gradually as x increases from 0.25 to 1.

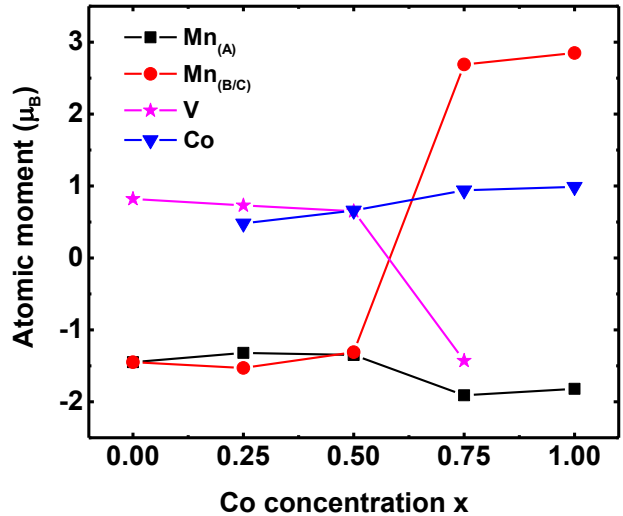

**Figure 6**: Variation of atomic moments with the Co concentration for $Mn_2V_{1-x}Co_xGa$.

L.Wollman *et al*. have carried out a detailed theoretical investigation on the structural and magnetic properties of the cubic $Mn_2Y^{(3d/4d)}Z$ (Z=Ga, Al) alloys [21]. The results show that for these Mn-based alloys with $Z_t \leq 24$, $L2_1$ structure is favoured (eg: $Mn_2VGa$) and when $Z_t > 24$, $X_a$ structure is favoured (eg: $Mn_2CoGa$). Another interesting observation is that due to the differences in the nearest neighbour atoms, $L2_1$ Mn-based alloys have ferromagnetic coupling for the Mn atoms at the A and C sub-lattices and in the case of $X_a$ Mn-based alloys, the coupling between the Mn atoms in the A and B sub-lattices is ferrimagnetic. This is one of the characteristic features of $X_a$ Mn-based alloys. This observation is comparable to our present investigation. In our investigation, the valence electron number ($Z_t$) of Mn$_2$VGa is varied from 22 to 26 by substituting Co at the V site. Our calculation also shows a ferromagnetic coupling between the Mn atoms at different sub-lattices for the alloys with x=0, 0.25 and 0.5 which crystallizes in the disordered $L2_1$ structure. The alloys with x=0.75 and 1 crystallize in the $X_a$ structure and the Mn moments are coupled ferrimagnetically.

Since the Mn$_2$V$_{0.5}$Co$_{0.5}$Ga alloy exhibits a fully compensated ferrimagnetic state with high T$_C$, it is interesting to investigate the half-metallic characteristic of Mn$_2$V$_{1-x}$Co$_x$Ga. Figure 7(a)-(e) shows the spin-resolved total density of states (TDOS) and partial density of states versus energy diagram for the Mn$_2$V$_{1-x}$Co$_x$Ga alloys. Only the dominant d band contribution of Mn, V and Co is considered for the partial density of states. As shown in figure 7(a), Mn$_2$VGa (x=0) exhibit a gap at the Fermi level in the spin-down subband. The non-zero density of states at the Fermi level in the spin-up subband is mostly contributed by the Mn$_{(A,C)}$ - 3d states. For Mn$_2$VGa, the calculated spin polarization is 92 % which indicates that it is nearly a half-metal. This is in agreement with the earlier reported values which is around 94 % [21,23]. For the Co substituted alloys also, the spin polarization values are more than 90 %. The calculated spin polarization values for the alloys with x=0.25, 0.5, 0.75 and 1 are 99, 99, 98 and 99 % respectively. As expected the V density of states decreases (shown in pink colour) and the Co density of states increases (shown in green colour) with the increase in Co concentration. The earlier *ab initio* results on the Mn$_{2-x}$Co$_x$VGa alloys (Co substitution at the Mn site) showed that the spin polarization decreases to 34 % for the compensation point x=1 [23]. The experimental investigation carried out by Ramesh Kumar *et al.* also indicates the absence of half-metallicity when Co is substituted at the Mn site [9]. In contrast to this, our present investigation shows that the spin polarization for Mn$_2$V$_{0.5}$Co$_{0.5}$Ga alloy is 99 %. The difference in the density of states curve for the spin up and spin down subbands indicates that the exchange potential for the magnetic atoms are having opposite signs resulting in the ferrimagnetic state which is evident from the atomic moments shown in Table 3 [28].

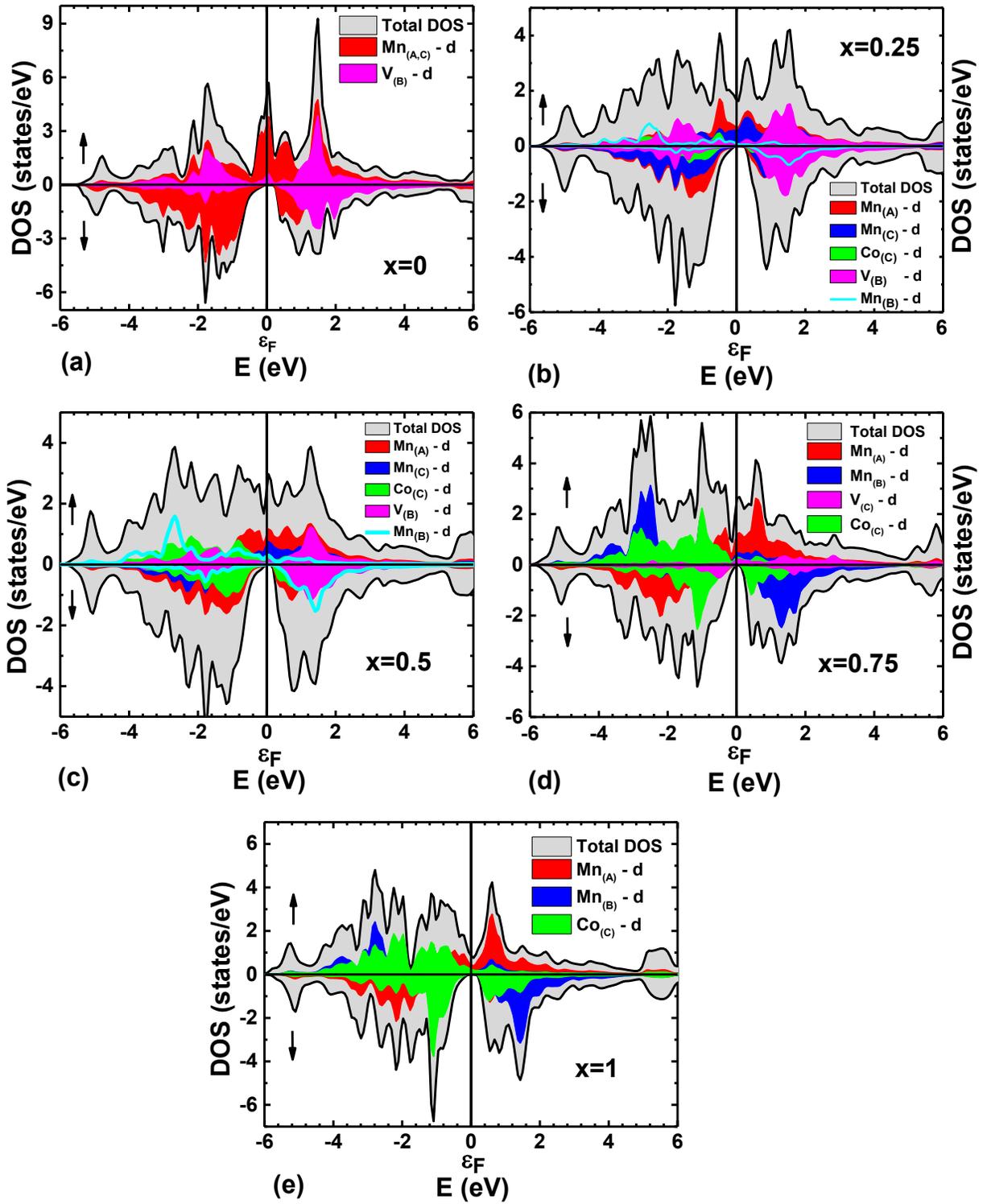

**Figure 7(a)-(e)**: Spin-resolved total and partial density of states versus energy diagram for the $Mn_2V_{1-x}Co_xGa$ (x=0, 0.25, 0.5, 0.75, 1) alloys.

## 4. Conclusions

The crystal structure and the magnetic configuration of $Mn_2V_{1-x}Co_xGa$ alloys which exhibit fully compensated ferrimagnetic property are identified through combined neutron diffraction and *ab initio* studies. Our investigation showed that the $L2_1$ and $X_a$ structure of the $Mn_2V_{1-x}Co_xGa$ alloys are well distinguishable from the neutron diffraction pattern which could not be identifiable from the X-ray diffraction studies. Neutron diffraction studies on $Mn_2V_{1-x}Co_xGa$ clearly indicated a structural transition from $L2_1$ to $X_a$ as the Co concentration increases. Alloy with x=0 crystallizes in $L2_1$, x=0.25 and 0.5 crystallizes in disordered $L2_1$, x=0.75 crystallizes in disordered $X_a$ and x=1 alloy crystallizes in $X_a$ structure. When a slight V-Ga disorder was present in $Mn_2VGa$, Mn-Co atomic antisite disorder was observed in the Co-substituted alloys. *Ab initio* studies have confirmed the structural transition observed in the neutron diffraction patterns. The identified magnetic configuration from the *ab initio* studies for x=0, 0.25, 0.5. 0.75 and 1 are **ddu**, **dduu, dduu, dudu** and **duu** respectively. It was evident that the disordered $L2_1$ structure with the antiparallel coupling between the ferromagnetically aligned ($Mn_{(A)}$-$Mn_{(C)}$) and (V-Co) atom pairs enables the magnetic moment compensation in $Mn_2V_{1-x}Co_xGa$. A high spin polarization value of more than 90 % was observed for all the alloys including the high $T_C$ fully compensated ferrimagnet $Mn_2V_{0.5}Co_{0.5}Ga$.

## Acknowledgement

P V Midhunlal acknowledge IIT Madras for all the funding and providing VIRGO supercluster for the High-Performance Computing Environment (HPCE). P V Midhunlal and N Harish Kumar acknowledge the UGC-DAE Consortium for Scientific Research, Mumbai Center, BARC for the neutron diffraction measurements.

# References

[1] Balke B, Fecher G H, Winterlik J and Felser C 2007 Mn$_3$Ga, a compensated ferrimagnet with high Curie temperature and low magnetic moment for spin torque transfer applications *Appl. Phys. Lett.* **90** 1–4.

[2] Hakimi M, Venkatesan M, Rode K, Ackland K and Coey J M D 2013 The zero-magnetization Heusler ferrimagnet *J. Appl. Phys.* **113** 2013–6.

[3] Galanakis I, Özdoğan K, Şaşinoğlu E and Aktaş B 2008 Ferrimagnetism and antiferro-magnetism in half-metallic Heusler alloys *Phys. Status Solidi Appl. Mater. Sci.* **205**, 1036–9.

[4] Akai H and Ogura M 2007 Hyperfine interactions of half-metallic diluted antiferromagnetic semiconductors *Hyperfine Interact.* **176** 21–5

[5] Lee K-W and Pickett W E 2008 Half semimetallic antiferromagnetism in the Sr$_2$CrTO$_6$ system (*T*=Os, Ru) *Phys. Rev. B* **77** 115101

[6] Long N H, Ogura M and Akai H 2009 New type of half-metallic antiferromagnet : transition metal pnictides *J. Phys.: Condens. Matter* **21** 064241

[7] Galanakis I, Dederichs P H and Papanikolaou N 2002 Slater-Pauling behavior and origin of the half-metallicity of the full-Heusler alloys *Phys. Rev. B - Condens. Matter Mater. Phys.* **66** 1–9

[8] Galanakis I, Özdoğan K, Şaşioğlu E and Aktaş B 2007 Doping of Mn$_2$VAl and Mn$_2$VSi Heusler alloys as a route to half-metallic antiferromagnetism *Phys. Rev. B - Condens. Matter Mater. Phys.* **75** 3–6

[9] Ramesh Kumar K, Arout Chelvane J, Markandeyulu G, Malik S K and Harish Kumar N 2010 Effect of Co substitution on the magnetic and transport properties of the half-metallic ferrimagnet Mn$_2$VGa *Solid State Commun.* **150** 70–3

[10] Deka B, Srinivasan A, Singh R K, Varaprasad B S D C S, Takahashi Y K and Hono K 2016 Effect of Co substitution for Mn on spin polarization and magnetic properties of ferrimagnetic Mn$_2$VAl *J. Alloys Compd.* **662** 510–5

[11] Ramesh Kumar K, Harish Kumar N, Markandeyulu G, Chelvane J A, Neu V and Babu P D 2008 Structural, magnetic and transport properties of half-metallic ferrimagnet Mn$_2$VGa *J. Magn. Magn. Mater.* **320** 2737–40

[12] Nagai K, Fujiwara H, Aratani H, Fujioka S, Yomosa H, Nakatani Y, Kiss T, Sekiyama A, Kuroda F, Fujii H, Oguchi T, Tanaka A, Miyawaki J, Harada Y, Takeda Y, Saitoh Y, Suga S and Umetsu R Y 2018 Electronic structure and magnetic properties of the half-metallic ferrimagnet Mn$_2$VAl probed by soft x-ray spectroscopies *Phys. Rev. B* **97** 035143 1–8

[13] Midhunlal P V, Chelvane J A, Krishnan U M A, Prabhu D, Gopalan R and Kumar N H 2018 Near total magnetic moment compensation with high Curie temperature Near total magnetic moment compensation with high Curie temperature in Mn$_2$V$_{0.5}$Co$_{0.5}$Z ( Z = Ga , Al ) Heusler alloys *J. Phys. D: Appl. Phys.* **51** 075002

[14] Siruguri V, Babu P D, Gupta M, Pimpale A V and Goyal P S 2008 A high resolution powder diffractometer *PRAMANA Journal of physics* **71** 1197–202

[15] Blaha P 2018 *WIEN2k* An Augmented PlaneWave Plus Local Orbitals Program for Calculating Crystal Properties vol 2

[16] Perdew J P, Burke K and Ernzerhof M 1996 Generalized Gradient Approximation Made Simple *Phys. Rev. Lett.* **77** 3865–8

[17] Anon 1994 Improved tetrahedron method for Brilleuin-zone integrations *Phys. Rev. B* **49**

[18] Kumar K R, Kumar N H, Babu P D, Venkatesh S and Ramakrishnan S 2012 Investigation of atomic anti-site disorder and ferrimagnetic order in the half-metallic Heusler alloy $Mn_2VGa$ *J. Phys. Condens. Matter* **24** 336007

[19] Liu G D, Dai X F, Liu H Y, Chen J L, Li Y X, Xiao G and Wu G H 2008 $Mn_2CoZ$ (Z = Al, Ga, In, Si, Ge, Sn, Sb) compounds : Structural, electronic, and magnetic properties 1–12 *Phys. Rev. B* **77** 014424

[20] Rodríguez-Carvajal J 1993 Recent advances in magnetic structure determination by neutron powder diffraction *Phys. B Condens. Matter* **192** 55–69

[21] Wollmann L, Chadov S, Kübler J and Felser C 2014 Magnetism in cubic manganese-rich Heusler compounds *Phys. Rev. B - Condens. Matter.* **90** 1–11

[22] Ozdogan K, Galanakis I, Şaşioglu E and Aktaş B 2006 Search for half-metallic ferrimagnetism in V-based Heusler alloys $Mn_2VZ$ (Z=Al, Ga, In, Si, Ge, Sn) *J. Phys. Condens. Matter* **18** 2905–14

[23] Li Q F, Zhao H F, Zhong X and Su J L 2012 Co doping effects on structural, electronic and magnetic properties in $Mn_2VGa$ *J. Magn. Magn. Mater.* **324** 1463–7

[24] Buschow K H J and van Engen P G 1981 Magnetic and magneto-optical properties of heusler alloys based on aluminium and gallium *J. Magn. Magn. Mater.* **25** 90–6

[25] Xing N, Li H, Dong J, Long R and Zhang C 2008 First-principle prediction of half-metallic ferrimagnetism of the Heusler alloys $Mn_2CoZ$ (Z = Al, Ga, Si, Ge) with a high-ordered structure *Comput. Mater. Sci.* **42** 600–5

[26] Alijani V, Winterlik J, Fecher G H and Felser C 2011 Tuning the magnetism of the Heusler alloys $Mn_{3-x}Co_xGa$ from soft and half-metallic to hard-magnetic for spin-transfer torque applications *Appl. Phys. Lett.* **99** 1–4

[27] Li G J, Liu E K, Zhang H G, Qian J F, Zhang H W, Chen J L, Wang W H and Wu G H 2012 Unusual lattice constant changes and tunable magnetic moment compensation in $Mn_{50-x}Co_{25}Ga_{25+x}$ alloys *Appl. Phys. Lett.* **101**

[28] Weht R and Pickett W 1999 Half-metallic ferrimagnetism in $Mn_2VAl$ *Phys. Rev. B* **60** 13006–10